# Wave-Encoded Model-based Deep Learning for Highly Accelerated Imaging with Joint Reconstruction

*Jaejin Cho[1,2], Borjan Gagoski[2,3], Taehyung Kim[1,2], Qiyuan Tian[1,2], Stephen Robert Frost[1,2], Itthi Chatnuntawech[4], and Berkin Bilgic[1,2,5]*

[1] Athinoula A. Martinos Center for Biomedical Imaging, Charlestown, MA, United States

[2] Harvard Medical School, Boston, MA, United States

[3] Boston Children's Hospital, Boston, MA, United States

[4] National Nanotechnology Center, Thailand

[5] Harvard/MIT Health Sciences and Technology, Cambridge, MA, United States

**\*Correspondence to**: Jaejin Cho, Ph.D.

jcho18@mgh.harvard.edu

**Running head:** Wave-encoded Model-based Deep Learning

# Abstract

**Purpose**: To propose a wave-encoded model-based deep learning (wave-MoDL) strategy for highly accelerated 3D imaging and joint multi-contrast image reconstruction, and further extend this to enable rapid quantitative imaging using an interleaved look-locker acquisition sequence with $T_2$ preparation pulse (3D-QALAS).

**Method**: Recently introduced MoDL technique successfully incorporates convolutional neural network (CNN)-based regularizers into physics-based parallel imaging reconstruction using a small number of network parameters. Wave-CAIPI is an emerging parallel imaging method that accelerates the imaging speed by employing sinusoidal gradients in the phase- and slice-encoding directions during the readout to take better advantage of 3D coil sensitivity profiles. In wave-MoDL, we propose to combine the wave-encoding strategy with unrolled network constraints to accelerate the acquisition speed while enforcing wave-encoded data consistency. We further extend wave-MoDL to reconstruct multi-contrast data with controlled aliasing in parallel imaging (CAIPI) sampling patterns to leverage similarity between multiple images to improve the reconstruction quality.

**Result**: Wave-MoDL enables a 47-second MPRAGE acquisition at 1 mm resolution at 16-fold acceleration. For quantitative imaging, wave-MoDL permits a 2-minute acquisition for $T_1$, $T_2$, and proton density mapping at 1 mm resolution at 12-fold acceleration, from which contrast weighted images can be synthesized as well.

**Conclusion**: Wave-MoDL allows rapid MR acquisition and high-fidelity image reconstruction and may facilitate clinical and neuroscientific applications by incorporating unrolled neural networks into wave-CAIPI reconstruction.

**Introduction**

To overcome the time-consuming scan of MRI, parallel imaging (PI) techniques have been developed to accelerate various MRI sequences through multiple receiver coils. Sensitivity encoding (SENSE (1)) and generalized auto-calibrating partial parallel acquisition (GRAPPA (2)) have been widely used to recover the image in the image domain and k-space, respectively, and the recent ESPIRiT (3) nicely bridged the sensitivity maps of SENSE and the interpolation kernel of GRAPPA. A number of techniques have been proposed to improve the conditioning of PI acquisition to push a higher acceleration. Controlled aliasing in parallel imaging results in higher acceleration (CAIPIRINHA (4)) modifies the appearance of aliasing artifacts during the acquisition to improve the subsequent PI reconstruction procedure for multi-slice imaging. Application of interslice shifts to three-dimensional (3D) imaging forms the basis of 2D-CAIPIRINHA (5), wherein the phase ($k_y$) and partition ($k_z$) encoding strategy is modified to shift the spatial aliasing pattern to reduce aliasing and better exploit the coil sensitivity variation.

Compressed sensing (CS) and low-rank reconstruction using annihilation filters further improved the MR reconstruction (6–9). In the last decade, PI combined with CS techniques has resulted in substantial improvements in acquisition speed and image quality. Although the PI-CS combination can achieve state-of-the-art performance (10–13), designing effective regularization schemes and tuning of hyper-parameters are non-trivial.

Wave-CAIPI is a more recent controlled aliasing method that can further reduce noise amplification and aliasing artifacts (14,15). It employs extra sinusoidal gradient modulations in the phase- and the partition-encoding directions during the readout to better harness coil sensitivity variations in all three dimensions. Wave encoding also incorporates 2D-CAIPI inter-slice shifts to improve the PI conditioning (5) and has found applications in highly accelerated gradient-echo (GE), MPRAGE, and fast spin-echo acquisitions (15–18).

Recently, image reconstruction with the deep learning approaches has been explored by several studies (19,20) to overcome the side effects of the existing reconstruction techniques such as long reconstruction time, noise amplification, and over-smoothness. A few studies have shown outstanding deep learning methods outperforming conventional regularization and/or optimization-based techniques in various applications. Line-by-line parallel imaging reconstruction has been implemented with multi-layer perceptron (19) and variational network (20) nicely incorporating the unrolled network during conjugate gradient updates. Some of the studies trained the neural networks in k-space to exploit the features in the spatial frequency domain (21–24) as many parallel imaging techniques have been implemented in k-space. In recent works, the wave-encoding strategy was successfully combined with the variational network to provide high-quality images at high acceleration (20,25).

Recently developed model-based deep learning (MoDL) improves image reconstruction by incorporating an unrolled CNN into parallel imaging forward model to help denoise and unalias undersampled data (26). MoDL consists of the unrolled CNN block and data consistency block. MoDL updates the image by iteratively passing both blocks. The sharing of network parameters across iterations enables MoDL to keep the number of learned parameters decoupled from the number of iterations, thus providing good convergence without increasing the number of trainable parameters. A smaller number of trainable parameters translates to significantly reduced training data demands, which is particularly attractive for data-scarce medical-imaging applications. The unrolled CNN block captures the image features to help the image reconstruction. The CNN will learn how to grab the image features from various reconstructed images during the iteration for image reconstruction. The data consistency block encourages consistency with measurements. At the data consistency block, an involving quadratic sub-problem was solved using conjugate gradient (CG) optimization. They used an alternating recursive algorithm to update the deep learning network. They further applied the MoDL approach to the multi-shot diffusion-weighted echo-planar MR imaging (27). They successfully replace the multi-shot sensitivity-encoded diffusion data recovery algorithm using structured low-rank matrix completion (MUSSELS) with the hybrid MoDL-MUSSELS which has two CNNs in both k-space and image domain. MoDL-MUSSELS could yield reconstructions that are comparable to state-of-the-art methods while offering several orders of magnitude reduction in run-time.

In this study, we proposed wave-encoded (wave-) MoDL for highly accelerated three-dimensional imaging. We incorporated wave-encoding strategy into MoDL reconstruction to better utilize sensitivity encoding that improves the reconstruction. We used the two unrolled CNNs in both image- and k-spaces that bring more efficient constraints. Furthermore, we extended wave-MoDL to reconstruct multi-contrast images using CAIPI sampling pattern to efficiently use multi-contrast information. We evaluated the improvement of multi-contrast image reconstruction on the multi-echo MPRAGE database for the human connectome project and the combination with 3D-quantification using an interleaved look–locker acquisition sequence with a T2 preparation pulse (3D-QALAS) database.

## Methods

- **Wave-CAIPI**

Wave-encoding employs additional sinusoidal gradients during the readout to take better advantage of 3D coil sensitivity profiles (14). Wave-encoding modulates the phase during the readout and incurs a corkscrew trajectory in k-space. The wave-encoded signal, $s$, can be explained using the following equation.

$$s(t) = \int_{x,y,z} m(x,y,z) e^{-2\pi i (k_x(t)x + k_y y + k_z z) - \gamma i \int_0^t (g_y(\tau)y + g_z(\tau)z) d\tau} dx dy dz$$

$$= \int_{x,y,z} m(x,y,z) e^{-2\pi i \left( k_x(t) \left( x + \frac{\gamma \int_0^t (g_y(\tau)y + g_z(\tau)z) d\tau}{2\pi k_x(t)} \right) + k_y y + k_z z \right)} dx dy dz \qquad [1]$$

where $m$ is the magnetization, $\gamma$ is the gyromagnetic ratio, and $g$ is the time-varying sinusoidal wave gradient. Voxel spreading by wave-encoding is a function of readout time and the position in the y- and z-dimensions, where is the $k_x$-y-z hybrid domain. The reconstruction of standard wave-CAIPI is described as follows,

$$x = \underset{x}{\mathrm{argmin}} \|\mathbf{W}\mathcal{F}_y\mathbf{P}\mathcal{F}_x\mathbf{C}x - b\|_2^2$$
$$= \underset{x}{\mathrm{argmin}} \|\mathbf{A}(x) - b\|_2^2 \qquad [2]$$

where $x$ is the reconstructed image, $\mathbf{W}$ is the subsampling mask, $\mathbf{P}$ is the wave point spread function in the $k_x$-y-z hybrid domain, $\mathbf{C}$ is the coil sensitivity map, and $b$ is the subsampled image, respectively.

- **Wave-MoDL**

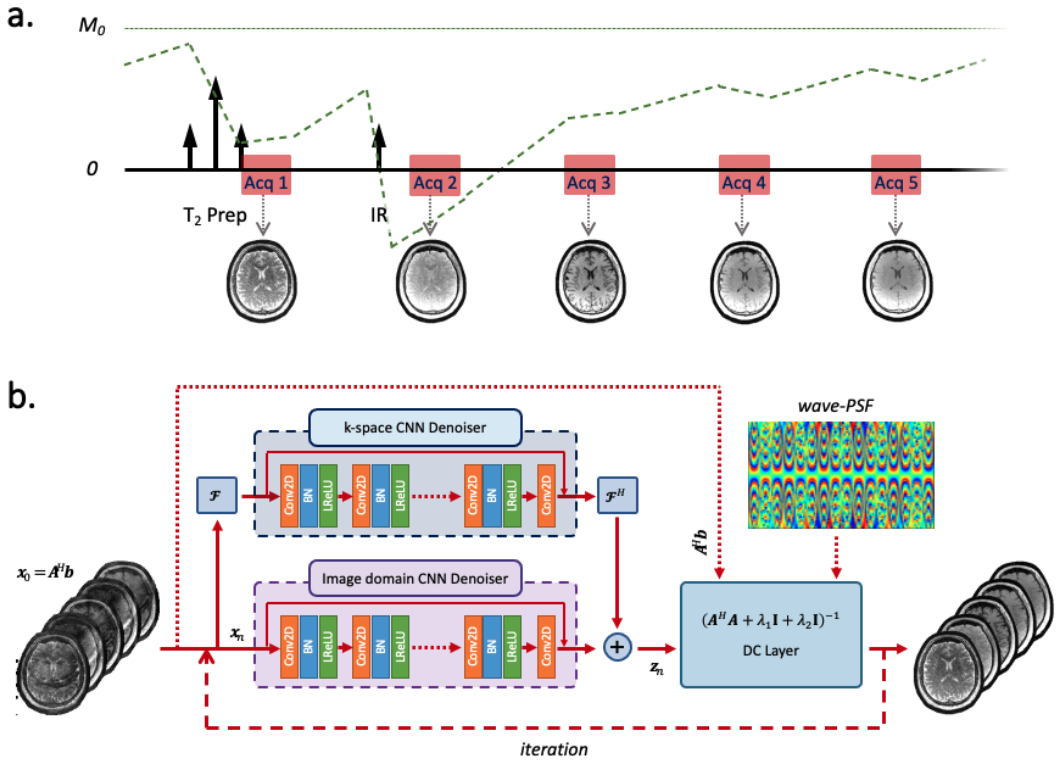

$$x = \underset{x}{\mathrm{argmin}} \sum_m^M \|\mathbf{W}_m\mathcal{F}_y\mathbf{P}\mathcal{F}_x\mathbf{C}x_m - b_m\|_2^2 + \lambda_1\|\mathbf{N}_k(x)\|_2^2 + \lambda_2\|\mathbf{N}_i(x)\|_2^2$$

**Figure 1.** a. 3D-QALAS sequence diagram. Five turbo-flash readout trains are played repeatedly until all k-space is acquired. The first readout succeeds a $T_2$-preparation module, and the remaining four readouts follow an inversion pulse that captures $T_1$ dynamics. b. Wave-MoDL diagram for multi-contrast joint reconstruction. $A = \mathbf{W}_m\mathcal{F}_y\mathbf{P}\mathcal{F}_x\mathbf{C}$ presents the forward model of wave-encoding. Convolutional regularizers are applied in both image- and k-space, which are combined with data consistency (DC) layers in an unrolled network structure. The model is trained in a supervised manner where the loss function measures the fidelity with respect to high-quality reconstructions at mild R=2-fold acceleration.

Figure 1b shows the proposed network architecture for wave-MoDL. We use the unrolled CNNs to constrain the reconstruction in both the image- and k-spaces, where several studies have reported that taking the networks in both domains improves the performance of the entire networks (21,27). The wave-MoDL reconstruction can be described as follows.

$$\begin{aligned} \boldsymbol{x} &= \operatorname*{argmin}_{\boldsymbol{x}} \|\mathbf{W}\mathcal{F}_y \mathbf{P}\mathcal{F}_x \mathbf{C}\boldsymbol{x} - \boldsymbol{b}\|_2^2 + \lambda_1 \|\mathbf{N}_k(\boldsymbol{x})\|_2^2 + \lambda_2 \|\mathbf{N}_i(\boldsymbol{x})\|_2^2 \\ &= \operatorname*{argmin}_{\boldsymbol{x}} \|\mathbf{A}(\boldsymbol{x}) - \boldsymbol{b}\|_2^2 + \lambda_1 \|\mathbf{N}_k(\boldsymbol{x})\|_2^2 + \lambda_2 \|\mathbf{N}_i(\boldsymbol{x})\|_2^2 \end{aligned} \quad [3]$$

where $\mathbf{N}_k$ and $\mathbf{N}_i$ represent residual CNNs in the k- and image-space, respectively. The residual CNN $\mathbf{N}$ can be explained by CNN $\mathbf{D}$ and the skip connection, $\mathbf{N}(\boldsymbol{x}) = \boldsymbol{x} - \mathbf{D}(\boldsymbol{x})$. We used the alternating minimization-based solution as described in (26,27), by which the network is unrolled. By substituting $\eta_{n+1} = \mathbf{D}_k(\boldsymbol{x}_{n+1})$ and $\zeta_{n+1} = \mathbf{D}_i(\boldsymbol{x}_{n+1})$, an alternating minimization-based solution to the problem iteratively.

$$\begin{aligned} \boldsymbol{x}_{n+1} &= (\mathbf{A}^H \mathbf{A} + \lambda_1 \mathbf{I} + \lambda_2 \mathbf{I})^{-1}(\mathbf{A}^H \boldsymbol{b} + \lambda_1 \eta_n + \lambda_1 \zeta_n) \\ \eta_{n+1} &= \mathbf{D}_k(\boldsymbol{x}_{n+1}) \\ \zeta_{n+1} &= \mathbf{D}_i(\boldsymbol{x}_{n+1}) \end{aligned} \quad [4]$$

- **Wave-MoDL for multi-contrast image reconstruction**

We further extended wave-MoDL to reconstruct multi-contrast images, $\boldsymbol{x}$, from the equation [2] and [3] as follows.

$$\begin{aligned} \boldsymbol{x} &= \operatorname*{argmin}_{\boldsymbol{x}} \sum_m^M \|\mathbf{W}_m \mathcal{F}_y \mathbf{P}\mathcal{F}_x \mathbf{C}\boldsymbol{x}_m - \boldsymbol{b}_m\|_2^2 + \lambda_1 \|\mathbf{N}_k(\boldsymbol{x})\|_2^2 + \lambda_2 \|\mathbf{N}_i(\boldsymbol{x})\|_2^2 \\ &= \operatorname*{argmin}_{\boldsymbol{x}} \sum_m^M \|\mathbf{A}_m(\boldsymbol{x}_m) - \boldsymbol{b}_m\|_2^2 + \lambda_1 \|\mathbf{N}_k(\boldsymbol{x})\|_2^2 + \lambda_2 \|\mathbf{N}_i(\boldsymbol{x})\|_2^2 \end{aligned} \quad [5]$$

where M is the number of image contrasts, $\mathbf{W}_m$ is the subsampling mask for m-th contrast, $\boldsymbol{x}_m$ is m-th contrast image, and $\boldsymbol{b}_m$ is the m-th subsampled image, respectively. CAIPI sampling patterns were applied across the contrasts to improve the reconstruction by efficiently using the data similarity of multi-contrast images.

- **Experiment**

We trained, validated, and tested the wave-MoDL using three different databases acquired on a 3T Siemens Prisma system with a 32-channel head coil. Table 1 shows the acquisition parameters of the databases and the networks. Figure 2 shows the g-factor analyses of SENSE and wave-CAIPI at the target acceleration for each database. Wave-

MoDL network updates the results during 10 iterations and takes 10 conjugate gradients per iteration in the data consistency layers. The unrolled CNN consists of 24-depth 5 hidden layers with leaky ReLU activation, and all network parameters were zero-initialized. Example code can be found at https://github.com/jaejin-cho/wave-modl.

Table 1. Parameters of database and network parameters

|  | MPRAGE | Multi-echo MPRAGE | QALAS |
|---|---|---|---|
| Imaging plane | Sagittal | Sagittal | Sagittal |
| Voxel size [mm$^3$] | 1 x 1 x 1 | 0.8 x 0.8 x 0.8 | 1 x 1 x 1 |
| FOV [mm$^3$] | 256 x 256 x 192 | 256 x 240 x 168 | 240 x 238 x 192 |
| Receiver bandwidth [Hz/pixel] | 200 | 744 | 347 |
| Maximum wave gradient [mT/m] | 8.80 | 9.63 | 16.51 |
| # of wave cycles | 11 | 4 | 5 |
| Target acceleration | 4 x 4 | 3 x 2 | 4 x 3 |
| # / depth of hidden layers | 5 / 24 | 5 / 24 | 5 / 24 |
| # of network parameters | 85,974 | 91,458 | 93,286 |
| # of subjects (train/validate/test) | 8 / 1 / 1 | 22 / 4 / 5 | 8 / 1 / 1 |

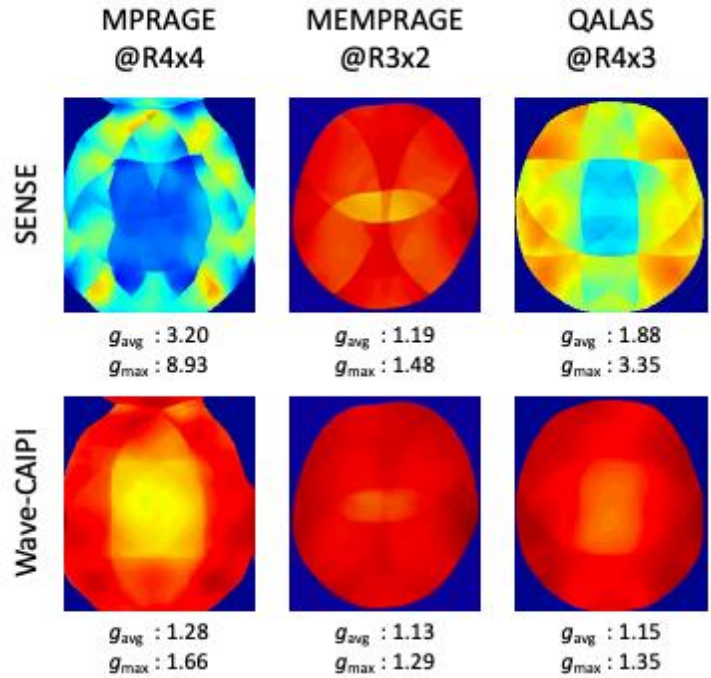

**Figure 2.** g-factor analyses of SENSE and wave-CAIPI at the target acceleration

- **MPRAGE dataset**

We scanned fully sampled T1-MPRAGE for healthy 10 subjects at 1-mm isotropic resolution to generate the database. We used 8 subjects, 1 subject, and 1 subject for training, validating, and testing the wave-MoDL network. We retrospectively subsampled the data to R=4x4, where wave encodings were applied in both $k_y$ and $k_z$ directions with 8.8 mT/m of $G_{max}$ and 11 cycles. The receiver bandwidth was 200 Hz/pixel.

- **Multi-echo MPRAGE dataset**

We used the four-echo MPRAGE database of the Human Connectome Project (HCP), consisting of 30 subjects at 0.8 mm isotropic voxel resolution and reduction factor 2. To make a clean database, we reconstructed images using the motion-corrected GRAPPA reconstruction for four different echo images separately. We retrospectively subsampled the data to R=3x2 and we incorporated the CAIPI sampling patterns to use complementary information for the reconstruction by using the data similarity across the multi echos. Wave encodings were applied in both $k_y$ and $k_z$ directions with 9.626 mT/m of $G_{max}$ and 4 cycles at 744 Hz/pixel of receiver bandwidth. Data used in the preparation of this work were obtained from the MGH-USC HCP database (https://ida.loni.usc.edu/login.jsp). The HCP project (Principal Investigators: Bruce Rosen, M.D., Ph.D., Martinos Center at Massachusetts General Hospital; Arthur W. Toga, Ph.D., University of Southern California, Van J. Weeden, MD, Martinos Center at Massachusetts General Hospital) is supported by the National Institute of Dental and Craniofacial Research (NIDCR), the National Institute of Mental Health (NIMH) and the National Institute of Neurological Disorders and Stroke (NINDS). Collectively, the HCP is the result of efforts of co-investigators from the University of Southern California, Martinos

Center for Biomedical Imaging at Massachusetts General Hospital (MGH), Washington University, and the University of Minnesota.

- **3D-QALAS dataset**

We scanned 3D QALAS for healthy 10 subjects at 1-mm isotropic resolution and reduction factor 2. The dataset includes 32-channel 3D-QALAS sagittal images acquired on a 3T Siemens Prisma scanner, consisting of five contrasts as shown in Figure 1a, for 10 subjects at 1mm-iso resolution. To reduce the chance of potential motion artifacts, data were acquired at R=2 to limit the scan time to 11 minutes. High-SNR reference images were reconstructed by GRAPPA (2) for five different contrast images separately. To train/validate/test the network, 8/1/1 subjects were used, respectively. The data were retrospectively subsampled to R=4x3, corresponding to a 1:52-minute quantitative exam, where 5-cycle cosine and sine wave-encodings were applied in both $k_y$ and $k_z$ directions with 16.5 mT/m of $G_{max}$ at 347 Hz/pixel bandwidth. CAIPI sampling patterns were employed to exploit complementary k-space information across the multiple contrasts. To take into account the signal intensity differences between the contrasts, the contrast images were weighted by [3.26,2.36,1.57,1.12,1], calculated by the signal norm ratio of each contrast image in the training dataset. This allowed each contrast to contribute to the loss function in similar amounts. A CAIPI sampling pattern across the multi contrasts was applied to multi-contrast MoDL and multi-contrast wave-MoDL to use the complementary information from other contrasts, while a fixed sampling pattern was applied to five different contrasts for SENSE and wave-CAIPI since these algorithms reconstruct each image independently.

**Results**

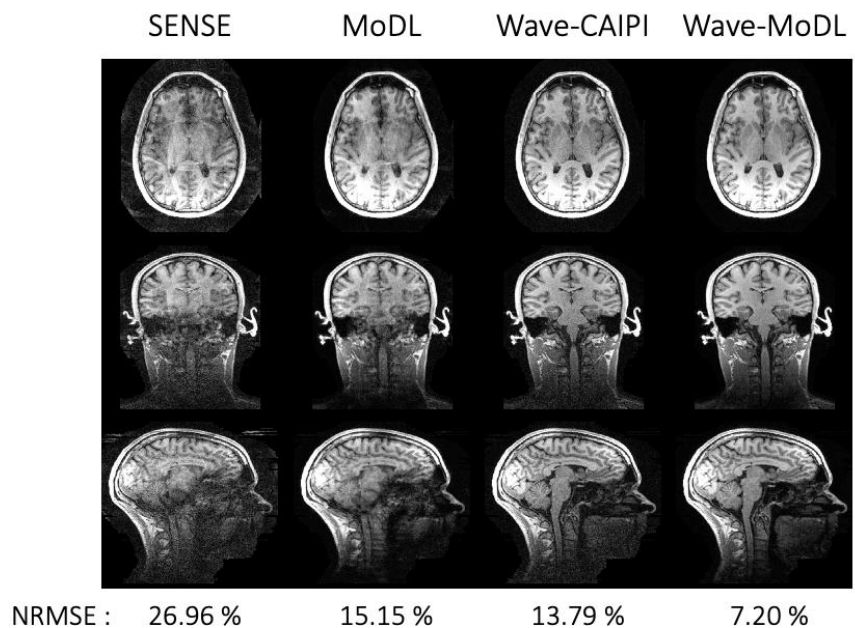

**Figure 3.** The proposed method on the MPRAGE database at R4x4

- **MPRAGE at R=4x4**

Figure 3 shows the reconstructed results at R=4x4 on the MPRAGE test set. SENSE suffers from the remaining folding artifact and noise amplification. MoDL significantly mitigated the noise amplification and reduce NRMSE by 1.78-fold, but still includes the folding artifact. Wave-CAIPI improved the reconstruction significantly, so NRMSE was reduced by 1.96-fold with respect to SENSE. Wave-MoDL further reduced NRMSE to 7.20%, which is the 1.92-fold improvement with respect to wave-CAIPI. Since wave-MoDL significantly improves the reconstruction at R=4x4, it allows acquiring whole-brain structural reference in 47 seconds at 1mm isotropic resolution.

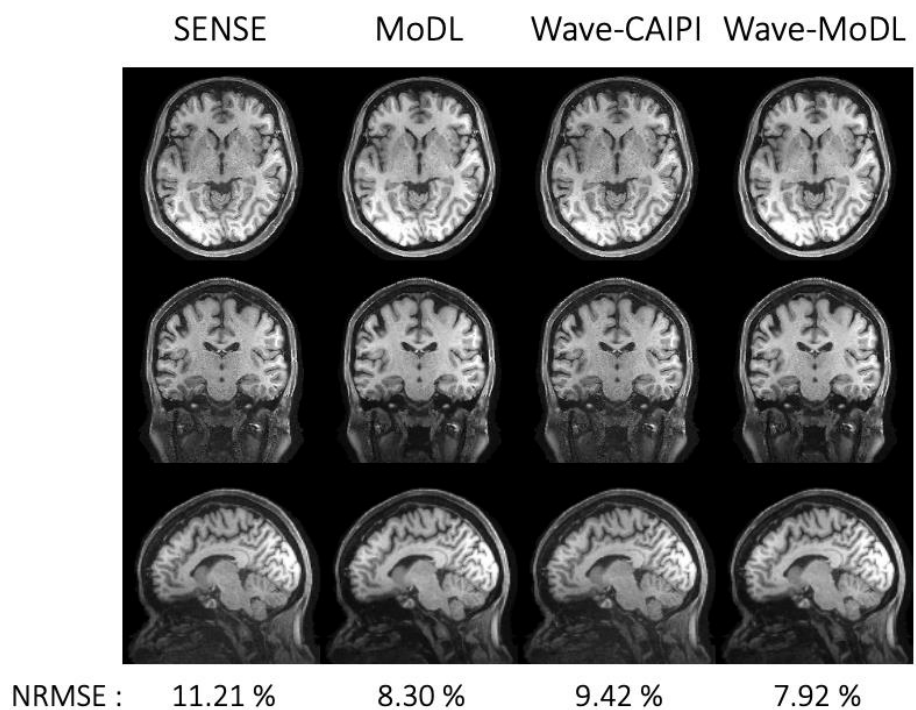

**Figure 4.** The proposed method on the MEMPRAGE database at R4x4. Echo images were combined with root mean squared.

- **Multi-echo MPRAGE at R=3x2**

Figure 4 shows the reconstructed images, the root-mean-squared images of four echoes, on the MEMPRAGE test set at R=3x2 to obtain sub-millimeter structural MRI. CAIPI sampling patterns across the multi echoes were applied to MoDL and wave-MoDL to fully use the multi-echo information to improve the reconstruction, while the fixed sampling patterns were applied to four different echoes for SENSE and wave-CAIPI reconstruction since these algorithms independently reconstruct each echo image and don't use the multi-echo information. Supporting Information Figure S1 shows the sampling patterns and each echo image before combination. SENSE suffers from the folding artifact and a little noise amplification. Although MoDL mitigated the noise amplification and some of the folding artifacts, it still includes folding artifacts as pointed by the arrows. Wave-encoding was pretty limited by a high receiver bandwidth (35), therefore wave-CAIPI also includes folding artifacts and even higher NRMSE with respect to MoDL. Wave-MoDL shows much improved reconstructed images and reduced NRMSE to 7.92%.

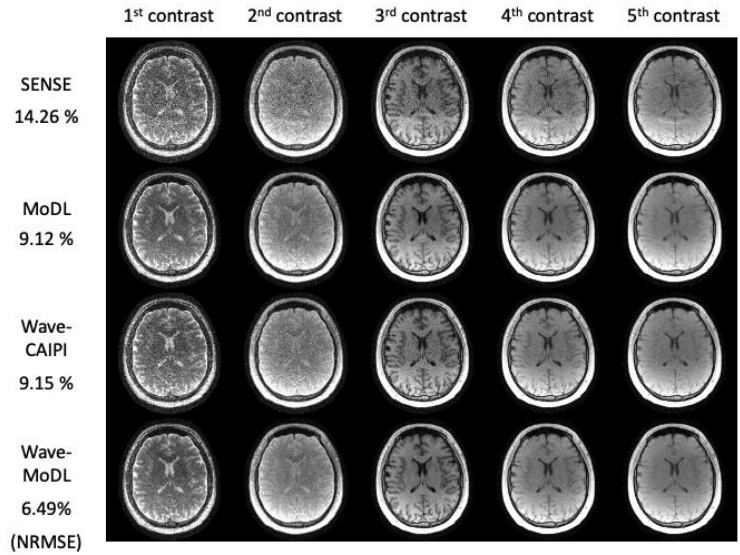

**Figure 5.** Multi-contrast image reconstruction using SENSE, MoDL, wave-CAIPI, and wave-MoDL.

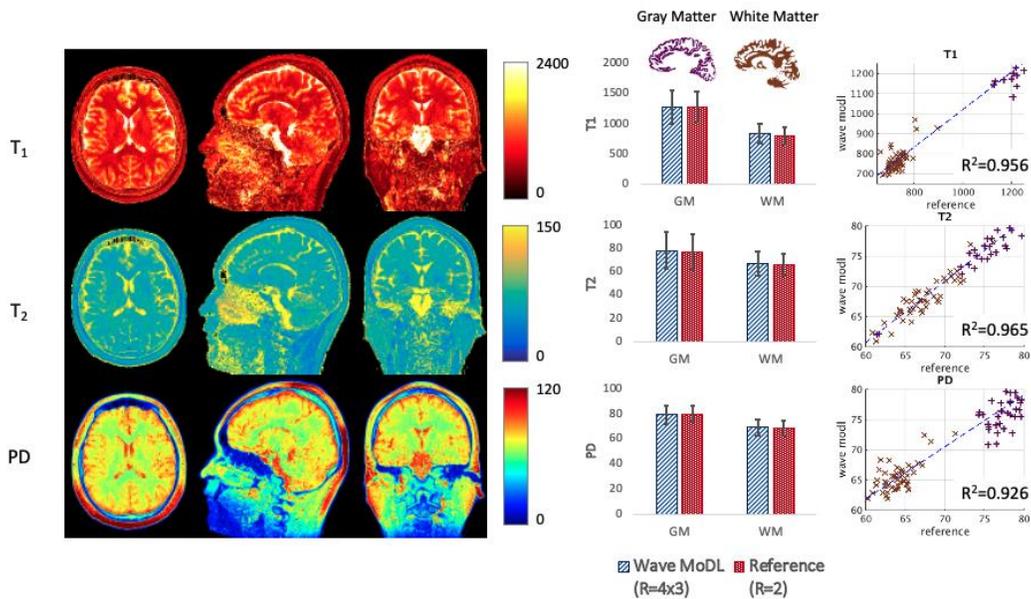

**Figure 6.** $T_1$, and $T_2$, PD maps were calculated using the wave-MoDL results. The last column shows the correction between the reference (R=2) and the quantitative values of wave-MoDL (R=4x3) in the randomly selected 5x5x5 boxes in gray matter (purple) and white matter (brown).

- **3D-QALAS at R=4x3**

Figure 5 shows the multi-contrast image reconstruction using SENSE, MoDL, wave-CAIPI, and wave-MoDL at R=4x3-fold acceleration. SENSE suffers from aliasing artifacts and noise amplification, while MoDL and wave-CAIPI mitigated noise amplification and reduced the NRMSE significantly. Incorporating wave-encoding improved the reconstruction and reduced the NRMSE to 9.15%. Wave-MoDL further reduced NRMSE to 6.49%, which is a 1.4-fold improvement with respect to wave-CAIPI. Figure 6 shows the $T_1$, $T_2$, and PD quantification results. To evaluate the accuracy, we calculated averaged values and standard deviation in white matter (WM, brown) and gray matter (GM, purple) segmented by FreeSurfer (33,34). In the last column, we selected 50 boxes that have 5x5x5 size in WM and GM and plotted the quantified values using the wave-MoDL results over the reference R=2 acquisition. The plotted graphs demonstrate that the estimated $T_1$, $T_2$, and PD values are well aligned with the reference. As shown in Figure 5, wave-MoDL much mitigated the noise amplification in the parameter maps as well.

**Discussion**

We introduced a wave-MoDL reconstruction incorporating wave-encoding that markedly improved image quality at extremely accelerated MR scan. In-vivo experiments show its ability to acquire the structural reference by 47-second MPRAGE, the sub-millimeter imaging with high fidelity by multi-echo MPRAGE, and quantitative imaging by 2.5-minute 3D-QALAS. We separated the 3D data into slice groups and trained on sets of aliasing slices. Though this approach is not able to use the information from adjacent slices, it significantly reduces the memory footprint and facilitates training with high channel count data.

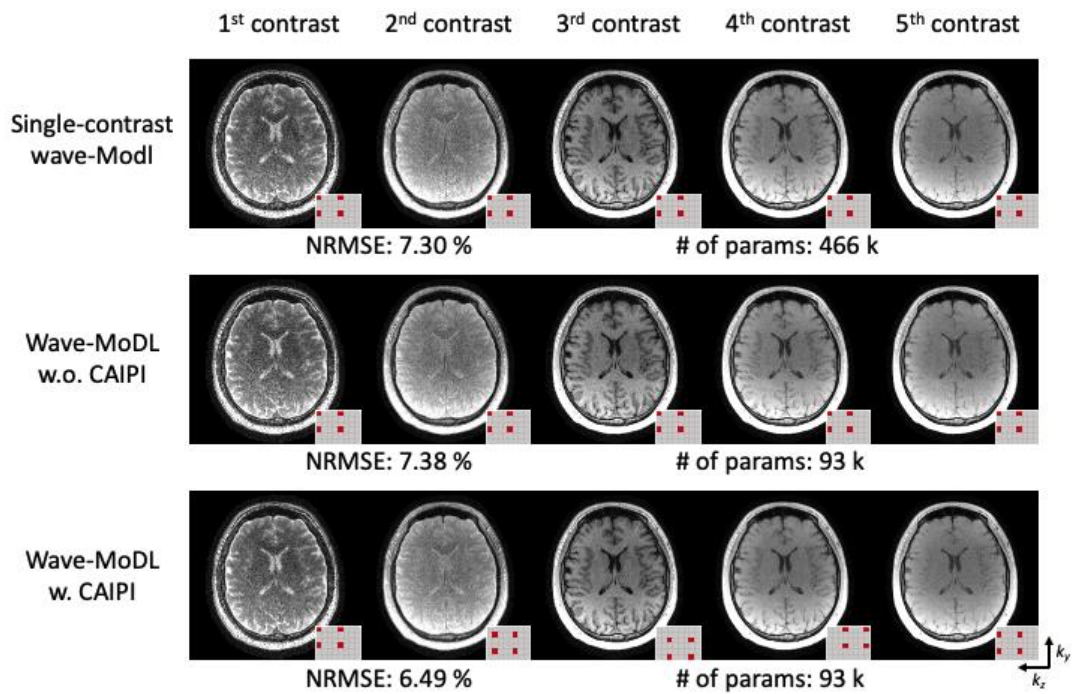

**Figure 7.** (top) Five individual single-contrast wave-MoDL reconstructions for each contrast and (middle and bottom) joint wave-MoDL with and without CAIPI sampling pattern across contrasts. The small pattern in the right bottom corner of each image shows the sampling in the $k_y$ and $k_z$ directions.

We employed the CAIPI sampling pattern across the multi echoes to improve the reconstruction by the similarity of the multi-echo images. To evaluate the efficiency of multi-echo information use, we reconstructed the images with and without the CAIPI sampling patterns on the QALAS database, furthermore, we trained five independent wave-MoDL networks to reconstruct each echo image separately, as shown in Figure 7. Figure 7 demonstrates that the use of CAIPI sampling patterns helps a lot to reconstruct the image by using multi-echo information and improves 0.81% of NRMSE. Wave-MoDL for multi-echo reconstruction without the CAIPI sampling patterns has comparable NRMSE with respect to independently trained networks for each echo while saving the number of network parameters by 5-fold. Wave-MoDL with the CAIPI sampling patterns has improved reconstructed images while saving the number of network parameters because the network can obtain complementary information from each other echo images.

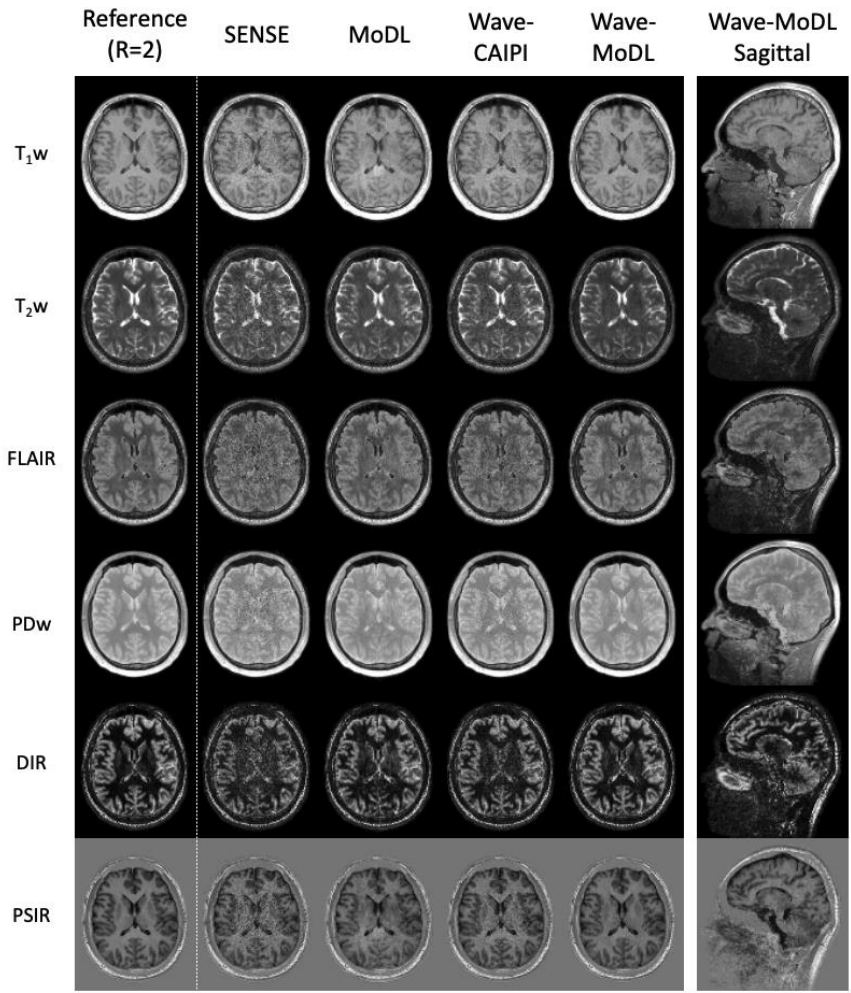

**Figure 8.** The synthesized $T_1$w, $T_2$w, FLAIR, PDw, double inversion recovery (DIR), and phase-sensitive inversion recovery (PSIR) images at R=4x3-fold acceleration.

3D-QALAS can estimate the $T_1$, $T_2$, and proton density parameter maps, thus every image contrast can be synthesized from the quantitative parameter maps. Figure 8 shows the synthesized $T_1$w, $T_2$w, FLAIR, PDw, double inversion recovery (DIR), and phase-sensitive inversion recovery (PSIR) images using the quantified $T_1$, $T_2$, and PD maps. The results demonstrate that wave-MoDL can mitigate the noise amplification and provide synthetic images that match well to the reference, compared with SENSE, MoDL, and wave-CAIPI.

## Conclusion

Wave-MoDL was proposed to enable highly accelerated MR scan to improve the geometric fidelity and acquisition speed. Its application was demonstrated in MPRAGE, multi-echo MPRAGE, and 3D-QALAS. Multi-echo wave-MoDL was introduced to improve the reconstruction by employing the information from other echoes for free with the CAIPI sampling patterns. In vivo experiments show the wave-MoDL acquisition allows high acceleration rates with decreased image artifacts relative to SENSE, MoDL, and wave-CAIPI.

## Acknowledgment

This work was supported by research grants NIH R01 EB028797, U01 EB025162, P41 EB030006, U01 EB026996, and R03EB031175 and the NVidia Corporation for computing support. Data collection and sharing for this project was provided by the MGH-USC Human Connectome Project (HCP; Principal Investigators: Bruce Rosen, M.D., Ph.D., Arthur W. Toga, Ph.D., Van J. Weeden, MD). HCP funding was provided by the National Institute of Dental and Craniofacial Research (NIDCR), the National Institute of Mental Health (NIMH), and the National Institute of Neurological Disorders and Stroke (NINDS). HCP data are disseminated by the Laboratory of Neuro Imaging at the University of Southern California.